\documentclass{article}

\usepackage{arxiv}

\usepackage[utf8]{inputenc} 
\usepackage[T1]{fontenc}    
\usepackage{hyperref}       
\usepackage{url}            
\usepackage{booktabs}       
\usepackage{amsmath}        
\usepackage{amsfonts}       
\usepackage{nicefrac}       
\usepackage{microtype}      
\usepackage{cleveref}       
\usepackage{lipsum}         
\usepackage{graphicx}
\usepackage{natbib}
\usepackage{doi}

\title{Modelling Cascading Physical Climate Risk in Supply Chains with Adaptive Firms: A Spatial Agent-Based Framework}

\author{ \href{https://orcid.org/0000-0002-4292-2367}{\includegraphics[scale=0.06]{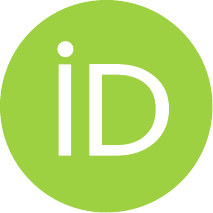}\hspace{1mm}Yara Mohajerani}\\
	 Quantile Labs\thanks{\url{https://www.quantile-labs.com}}\\
	Toronto, Ontario, Canada \\
	\texttt{yara@quantile-labs.com} \\
}

\renewcommand{\headeright}{}
\renewcommand{\undertitle}{}
\renewcommand{\shorttitle}{Modelling Cascading Physical Climate Risk in Supply Chains with Adaptive Firms: A Spatial Agent-Based Framework}

\hypersetup{
pdftitle={Modelling Cascading Physical Climate Risk in Supply Chains with Adaptive Firms: A Spatial Agent-Based Framework},
pdfsubject={cs.MA, q-fin.RM},
pdfauthor={Yara Mohajerani},
pdfkeywords={Agent-based modelling \and Physical climate risk \and Supply-chain resilience \and Cascading disruptions \and Climate adaptation \and Scenario analysis},
}

\begin{document}

\maketitle

\begin{abstract}
We present an open-source Python framework for modelling cascading physical climate risk in a spatial supply-chain economy. The framework integrates geospatial flood hazards with an agent-based model of firms and households, enabling simulation of both direct asset losses and indirect disruptions propagated through economic networks. Firms adapt endogenously through two channels: capital hardening, which reduces direct damage, and backup-supplier search, which mitigates input disruptions. In an illustrative global network, capital hardening reduces direct losses by 26\%, while backup-supplier search reduces supplier disruption by 48\%, with both partially stabilizing production and consumption. Notably, firms that are never directly flooded still bear a substantial share of disruption, highlighting the importance of indirect cascade effects. The framework provides a reproducible platform for analyzing systemic physical climate risk and adaptation in economic networks.
\end{abstract}

\section{Introduction}
\label{sec:intro}

Climate change poses risks to the economic system through physical and transition risks \citep{batten2016let,monasterolo2020climate,ranger2022assessing}. Transition risks arise from the shift toward a greener economy, such as higher carbon taxes and their downstream effects through production and finance. Physical risks arise from the changing state of the climate itself, through acute hazards such as floods and wildfires as well as chronic hazards such as rising temperatures and sea-level rise \citep{ranger2022assessing}. Chronic risks are typically assessed with aggregate economic damage functions, such as those of \citet{kalkuhl2020impact} and \citet{kotz2024economic}. In institutional climate scenario analysis, acute damages are often quantified at asset level using structural damage functions. However, direct asset damages alone do not capture the systemic risks created when supply chains, labour markets, and firm finances transmit shocks beyond the directly exposed locations.

Existing econometric damage functions show a wide range of estimates due to varying assumptions and modelling limitations. According to NGFS v5 reconstructed damage functions, effects on GDP by 2100 range from 2\% to 44\% under the NGFS Current Policies scenario \citep{NGFS2024DamageFunctions}. These limitations include extrapolation beyond historically observed climate regimes \citep{rising2022challenges,franzke2022perspectives}, restrictive functional-form assumptions \citep{burke2015global,kotz2024economic,neal2025reconsidering}, and limited treatment of nonlinear cascades through socioeconomic networks \citep{franzke2022perspectives,rising2022challenges}. These limitations are especially problematic for acute physical risk, where realized impacts depend not only on direct exposure but also on how agents adjust behaviour, reallocate spending, and propagate shortages through supply chains.

Agent-based modelling (ABM) offers a promising alternative by enabling bottom-up simulation of heterogeneous agents and their interactions \citep{farmer2009economy}. In the climate-risk literature, ABMs can propagate shocks through networks of firms and households, generating endogenous indirect effects that are invisible in direct-damage-only approaches \citep{lamperti2019towards}. Several prior frameworks have examined disaster impacts on production networks using disaggregated firm-level input-output models \citep{henriet2012firm,bierkandt2014acclimate,wenz2014acclimate}. These studies demonstrate the value of network structure for disaster propagation, but they typically rely on fixed parametric responses, short post-disaster horizons, or limited software support for long-run scenario workflows.

A recent survey by the Bank of England found that most ABMs used for climate risk quantification by central banks still focus on transition risk rather than physical risk \citep{borsos2025agent}. A notable exception is \citet{gourdel2023non}, which incorporated both physical and transition risk into a stress-testing framework for investment funds, but on a much shorter time horizon than the century-scale scenario analysis typically used in climate risk assessment. A persistent software challenge therefore remains: how to integrate geospatial climate hazard data, interpretable firm behaviour, reproducible scenario configuration, and systemic-risk diagnostics into a single open framework that is practical for long-horizon physical-risk analysis.

We address this gap by developing an open-source spatial ABM framework for acute physical climate risk. The framework combines bottom-up asset-level damage functions, long-run physical hazard scenarios, a demand-driven household-firm economy, and an explicit hazard-conditional continuity-adaptation module. Three aspects are central. First, firms are geolocated directly on a high-resolution hazard grid and linked through a supply-chain network. Second, core operating rules are fixed and transparent, while climate adaptation is isolated in a dedicated continuity-capacity stock that responds to hazard-induced operating shortfall and deploys through one of two interpretable channels: backup-supplier search or capital hardening. The shared stock is a reduced-form preparedness resource used as an experimental control, so the comparison isolates deployment-channel differences under a common perception and funding rule rather than claiming identical micro-level costs or organizational processes. Third, the software workflow is built for reproducible scenario analysis, with parameter-file configuration, matched-seed ensembles, self-describing output metadata, and diagnostics for indirect cascade risk on firms that remain never directly flooded. We demonstrate the framework on a 100-firm illustrative network under RCP8.5 riverine flooding from 2020 Q1 to 2099 Q4; this application is intended as an internal evaluation of the framework's behaviour rather than as an empirically calibrated case study of a specific real economy.

The contribution of this paper is therefore both methodological and software-oriented. Methodologically, it provides an ABM architecture for distinguishing direct damage mitigation from indirect continuity preservation. From a software perspective, it provides an openly available workflow with self-describing outputs for configuring, running, extending, and diagnosing physical climate risk experiments under repeated hazards. The paper evaluates that workflow through internal consistency, warm-up, dormancy, and sensitivity checks, not through external calibration or historical backtesting.

\section{Framework Design and Implementation}
\label{sec:methodology}

The framework is implemented in Python using the Mesa agent-based modelling framework \citep{kazil2020utilizing} for agent scheduling, spatial grid management, and data collection. The codebase separates model orchestration, agent behaviour, command-line execution, ensemble utilities, sensitivity analysis, and visualization into distinct modules. Scenario configuration is primarily driven by JSON parameter files, with command-line overrides for common experimental controls such as hazard toggles, adaptation strategy, continuity-sensitivity bounds, and seed lists.

\subsection{Software Architecture and Reproducibility Workflow}
\label{sec:software}

The main runtime components are \texttt{model.py}, which defines the economy-wide scheduler, hazard integration, and data collection; \texttt{agents.py}, which defines household and firm behaviour; and \texttt{run\_simulation.py}, which exposes the scenario workflow through a command-line interface. Supporting modules handle hazard sampling, ensemble merging, matched-seed sensitivity analysis, and figure generation. This separation keeps the policy-relevant model logic visible while allowing scenario orchestration and post-processing to evolve independently.

Reproducibility is treated as a first-class software requirement. Multi-seed runs can be executed in one command or extended later by merging member panels. Summary and member CSV outputs include \texttt{Meta\_*} metadata fields that record the effective scenario label, parameter file, topology file, hazard schedule, seed range, adaptation settings, and explicit command-line overrides. This makes each output file self-describing even when adaptation choices are not hard-coded in the underlying JSON parameter file. The plotting scripts consume these summary files directly and, when dispersion bands are required, can read the matching member sidecars so that ensemble bands are computed from the correct member trajectories rather than reconstructed from already-aggregated statistics. In this paper, the same workflow also supports the reported internal evaluation checks by making accounting diagnostics, control scenarios, and ensemble diagnostics directly observable.

\subsection{Network Setup}
\label{sec:network}

\begin{figure}[!t]
\centering
\includegraphics[width=0.95\textwidth]{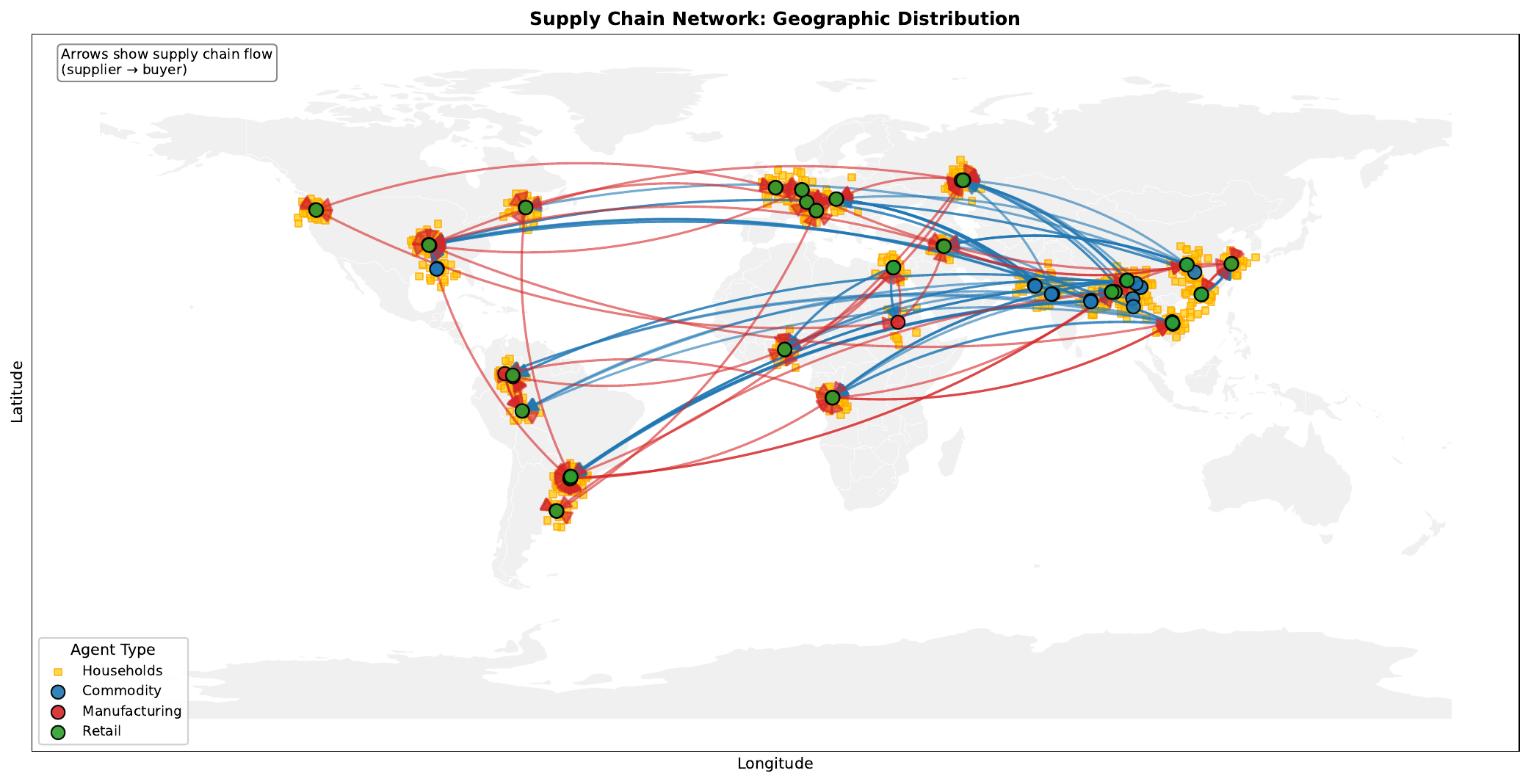}
\caption{The geographical distribution of the illustrative economy of 100 global firms, consisting of 30 commodity firms (blue), 45 manufacturing firms (red), and 25 retail firms (green). Arrows indicate the supply-chain flow of goods. Interactions with the 1000 households providing labour to and buying goods from firms are not shown. Map lines delineate study areas and do not necessarily depict accepted national boundaries.}
\label{fig:geography}
\end{figure}

We construct a network of economic agents on a 0.25-degree resolution global grid (1440$\times$720 cells). A simplified economic network is utilized in this study, with 100 firms placed onto the grid around flood-prone locations. In addition, 1000 households are initialized on the grid, in close vicinity to the firms, giving a 10:1 household-to-firm ratio. The distribution of the firm agents is represented in Figure \ref{fig:geography}.

The network of firms is composed of commodity, manufacturing, and retail firms represented by nodes, with directed edges denoting supplier-buyer relationships. Commodity firms occupy the upstream tier with no incoming edges; manufacturing firms receive inputs from commodity producers; retail firms receive inputs from manufacturers and act as the final-consumption interface to households. Household demand is applied only to final-good sectors; in the presented sample network topology, retail is the only final-good sector, so the effective household consumption ratio is 100\% retail. This three-tier structure (commodity $\rightarrow$ manufacturing $\rightarrow$ retail $\rightarrow$ households) creates idealized propagation pathways for both supply shocks (upstream disruptions cascade downstream) and demand shocks (reduced consumption propagates upstream through reduced orders).

Firms also trade within the same sector, so sector membership does not uniquely determine network position even within the simplified three-tier layout.

\subsection{Acute Risk Application}
\label{sec:hazard}

This study focuses on riverine flood risk under RCP8.5 as an illustration of acute flooding shocks on the economic network. Riverine flood depth projections are obtained from the World Resources Institute Aqueduct database \citep{ward2020aqueduct}. The original data is resampled to the same 0.25-degree grid as the spatial agent model using mean aggregation, which represents the average hazard exposure across each cell rather than the worst-case pixel. Flood depths are sampled independently for each grid cell based on return-period frequencies.

Damages are calculated using JRC Global Flood Depth-Damage Functions \citep{huizinga2017global}. The sector and location of each agent determine the corresponding flood depth versus damage curve. Sector mapping assigns industrial damage curves to commodity and manufacturing firms, and commercial curves to retail firms. The JRC curves provide a direct loss fraction for each hazard realization. This loss fraction affects agents through three pathways: (1) firm capital stock reduction, (2) temporary productivity losses through an evolving productivity state, and (3) inventory destruction. The productivity state is updated downward when new losses occur and then recovers gradually over time. Recovery of this productivity state is liquidity-dependent: the recovery rate scales from 20\% per step for firms with near-zero cash to 50\% per step for firms with ample liquidity (money $\geq 200$ monetary units). Initial inventories, working capital, and installed capital are seeded from a demand-consistent bootstrap that propagates final demand upstream through the supplier network and scales the implied activity to the available household labour force.

\subsection{Economic Agents}
\label{sec:agents}

\subsubsection{Households}
\label{sec:households}

Each household agent supplies one unit of labour per simulation step. Employment decisions use a staged search. Households first rank nearby same-sector firms, and only if none of those firms hire them do they broaden the search to the wider market with an additional remote-search penalty. Candidate employers are scored using a wage-distance utility augmented by sector-match and fallback-search terms:
\begin{equation}
\label{eq:utility}
U_{hf} = w_f - \delta_h \cdot d_{hf}
+ \beta_{\text{match}} \mathbf{1}(s_f = s_h)
- \beta_{\text{mismatch}} \mathbf{1}(s_f \neq s_h)
- \beta_{\text{remote}} \mathbf{1}(f \notin \mathcal{N}_h)
\end{equation}
where $w_f$ is the wage offered by firm $f$, $d_{hf}$ is the Manhattan distance between household $h$ and firm $f$, $\delta_h$ is a household-specific distance cost parameter, $s_f$ and $s_h$ are firm and household sectors, and $\mathcal{N}_h$ is the nearby candidate set used in the first search stage. The fixed scoring terms are $\beta_{\text{match}} = 0.15$, $\beta_{\text{mismatch}} = 0.20$, and $\beta_{\text{remote}} = 0.10$. The distance cost parameter is randomized per household from a uniform distribution from 0.01 to 0.1 to create heterogeneous preferences while modelling widespread availability of remote work. Cross-sector employment remains permitted through the second-stage fallback search, but the primary search is sector-local so that upstream production is not systematically starved by globally higher-paying downstream firms.

Household consumption follows a two-stage budgeting process. Households are treated as the residual owners of firms, so their disposable income includes both labour earnings and firm payouts. First, the consumption budget is determined as:
\begin{equation}
\label{eq:budget}
B_c = \eta_Y \left(Y_h^L + Y_{h,t-1}^D + Y_{h,t-1}^K + Y_{h,t-1}^A\right) + \eta_M \max(0, M_h - \bar{M})
\end{equation}
where $Y_h^L$ is current-period labour income, $Y_{h,t-1}^D$ is dividend income distributed after the previous completed period, $Y_{h,t-1}^K$ is reduced-form capital-service income returned when firms install productive capital (Section \ref{sec:budgeting}), $Y_{h,t-1}^A$ is reduced-form adaptation-service income returned when firms fund continuity maintenance or investment (Section \ref{sec:adaptation}), and $M_h$ is household money holdings. In the current closed-economy closure, these non-wage firm payouts are pooled and distributed evenly across households rather than tied to household-specific ownership claims. We set $\eta_Y = 0.9$ so that most current income is recycled into demand, $\eta_M = 0.02$ so that wealth effects remain modest, and $\bar{M}=50$ as a precautionary cash target equal to half of the initial household money endowment. The budget is capped at current money holdings so households cannot spend more liquidity than they possess.

Second, the budget is allocated across final-good sectors according to consumption ratios $\rho_s$. In the current 100-firm network topology, retail is the only final-good sector, so the effective allocation is 100\% retail. Within each eligible sector, households purchase from firms in ascending price order until the sector budget is exhausted, with fractional purchases permitted when budget is less than a unit price. Consumption occurs after current-period production, so households buy from goods made available within the same simulation period rather than from a purely lagged inventory stock.

\subsubsection{Firm Production}
\label{sec:production}

Firms produce output using a Leontief production function \citep{leontief1986input} with three inputs: labour $L$, intermediate inputs $I$, and capital $K$, with realized output also capped by the demand-driven target $Q^*$ determined in the planning stage:
\begin{equation}
\label{eq:production}
Q = \min\left(Q^*, \phi \cdot \min\left(\frac{L}{\alpha_L^s}, \frac{I}{\alpha_I^s}, \frac{K}{\alpha_K^s}\right)\right)
\end{equation}
where $\alpha_L^s$, $\alpha_I^s$, and $\alpha_K^s$ are sector-specific production coefficients representing input requirements per unit output, $Q^*$ is the planned demand-driven output target from Section \ref{sec:budgeting}, and $\phi \in [0,1]$ is the firm's remaining productivity state after accumulated direct damage and partial recovery.

\begin{table}[!t]
\centering
\begin{tabular}{lccc p{5.2cm}}
\toprule
\textbf{Sector} & $\alpha_L$ & $\alpha_I$ & $\alpha_K$ & \textbf{Rationale} \\
\midrule
Commodity & 0.6 & 0.0 & 0.7 & Upstream boundary condition with the highest capital intensity \\
Manufacturing & 0.3 & 0.6 & 0.6 & Highest intermediate-input requirement in the simplified topology \\
Retail & 0.5 & 0.4 & 0.2 & Lower capital and intermediate-input requirement \\
\bottomrule
\end{tabular}
\caption{Sector-specific Leontief technical coefficients. Lower values indicate higher productivity.}
\label{tab:coefficients}
\end{table}

Table \ref{tab:coefficients} reports the sector-specific production coefficients. The coefficients are stylized but chosen to preserve broad sector ordering consistent with OECD structural-analysis evidence \citep{horvat2020oecd,korinek2020mining}: manufacturing is assigned the highest intermediate-input requirement, commodity production the highest capital intensity, and retail lower capital and intermediate-input requirements. The zero intermediate-input coefficient for commodity firms is a modelling boundary condition of the simplified three-tier topology rather than an empirical claim about extractive industries, which in practice also purchase substantial upstream services and other inputs \citep{korinek2020mining}. The coefficients should therefore be interpreted as empirically informed stylizations rather than a calibration to a specific national input-output table; future work using country-specific or firm-level data could replace them with directly calibrated values.

Inputs from multiple connected suppliers are treated as substitutable (sum-based aggregation), allowing production to continue even if individual suppliers face disruptions, provided total input availability meets requirements.

\subsubsection{Firm Planning and Operating Cycle}
\label{sec:budgeting}

Firms do not allocate fixed expenditure buckets ex ante. At the start of each period, each firm forms an expected-sales forecast from realized sales and sets an initial demand-driven target equal to expected sales plus a fixed inventory buffer ratio, net of current finished-goods inventory. The planned output target used in Equation \ref{eq:production} is therefore
\begin{equation}
\label{eq:qstar}
\begin{aligned}
Q^*_{i,t} = \min \Bigg(&\max\left(0, \hat{S}_{i,t} + B_{i,t} - O_{i,t}\right),\frac{K_{i,t}}{\alpha_K^s}\phi_{i,t},\frac{\mathcal{L}^{\mathrm{op}}_{i,t}}{c_{i,t}}\Bigg),
\end{aligned}
\end{equation}
where $\hat{S}_{i,t}$ is expected sales, $B_{i,t}$ is the finished-goods inventory target, $O_{i,t}$ is current finished-goods inventory, $\phi_{i,t}$ is the current productivity state, $\mathcal{L}^{\mathrm{op}}_{i,t}$ is available operating finance, and $c_{i,t}$ is the unit variable cost implied by current wages and input prices. To prevent firms from exhausting liquidity in one period while still allowing current production to respond to demand, each firm maintains a liquidity buffer equal to the larger of 10 monetary units and 15\% of current cash holdings. Payroll and intermediate-input purchases may additionally draw a bounded operating overdraft backed by the firm's recent or expected sales. Available operating finance is therefore the sum of cash above the liquidity buffer and this bounded working-capital facility.

Vacancy demand and intermediate-input demand are derived mechanically from desired output using the Leontief coefficients. Connected suppliers are treated as substitute sources of one aggregate intermediate good, and buyers procure from the cheapest available suppliers first until the required volume or the operating-finance limit is reached. Capital is represented as a reduced-form productive-capacity stock rather than as an explicit traded capital good. After payroll, intermediate purchases, and depreciation are accounted for, positive profits first rebuild the firm's base capital target, then support any remaining demand-driven expansion, and only thereafter flow to adaptation spending or household dividends. Because the model does not include a separate capital-goods-producing sector, the cash outlay associated with this reduced-form investment is returned to households as investment income so that the circular flow of money remains closed.

Each simulation period is phased. Firms first update adaptation decisions using the previously observed hazard state, but the associated spending is not deducted immediately. The new hazard state is then sampled and firms plan operations under that hazard realization; households supply labour; firms hire, procure inputs, and produce in broad sector-tier order (commodity before manufacturing before retail, with random tie-breaking within sector); households then consume final goods; and firms finally close sales and accounting records for the period. At this closing stage, maintenance and new continuity spending are funded from residual post-operations cash and therefore affect the next period's continuity-capacity stock rather than the current period's hiring and procurement. Aggregate household-plus-firm money is tracked each step as an accounting diagnostic to verify stock-flow closure.

\subsubsection{Wage Dynamics}
\label{sec:wages}

Wages are set through revenue-based targeting, in which each firm anchors its wage offer to the revenue generated per worker. The target wage is:
\begin{equation}
\label{eq:target_wage}
w^* = \frac{R}{L_{\text{hired}}} \cdot \lambda
\end{equation}
where $R$ is the firm's revenue from the previous completed period, $L_{\text{hired}}$ is the number of workers employed, and $\lambda = 0.5$ is a fixed labour share of revenue. Firms that had no workers in the previous period use a modest entry fallback equal to 1.02 times the current market mean wage, while firms with workers but zero revenue keep their previous wage offer.

Wages adjust smoothly toward the target at 10\% per step to damp quarter-to-quarter oscillation:
\begin{equation}
\label{eq:wage_adj}
w_{t+1} = w_t + 0.1 \cdot (w^* - w_t)
\end{equation}

This mechanism is self-correcting: wages are structurally bounded by what workers produce and automatically decrease when revenue falls. A stylized wage floor at 40\% of the initial mean wage prevents unrealistically deep wage collapse during prolonged hazard slumps.

\subsubsection{Price Dynamics}
\label{sec:prices}

Prices are set through a simple markup pricing mechanism anchored to unit costs. The rule is intended as a transparent demand-pressure closure rather than as a calibrated industrial-organization model. The unit cost is:
\begin{equation}
\label{eq:unit_cost}
c = \frac{\alpha_L^s \cdot w + \alpha_I^s \cdot \bar{p}_I}{\phi}
\end{equation}
where $w$ is the firm's wage, $\bar{p}_I$ is the average input price from suppliers, and $\phi$ is the current productivity state. The target price is:
\begin{equation}
\label{eq:target_price}
p^* = c \cdot (1 + m)
\end{equation}
where $m$ is the markup, determined by the sell-through rate $\sigma$ from the previous completed period. The implemented linear rule is
\begin{equation}
\label{eq:markup}
m = 0.02 + 0.30 \sigma,
\end{equation}
so the markup remains modest and always positive over $\sigma \in [0,1]$. Prices adjust smoothly toward the target at 20\% per step to damp oscillation:
\begin{equation}
\label{eq:price_adj}
p_{t+1} = p_t + 0.2 \cdot (p^* - p_t)
\end{equation}
with an absolute floor of 0.5 to prevent degenerate zero or negative prices.

Pricing is rule-based rather than directly adapted; continuity investment affects prices only indirectly through realized sales, input costs, inventories, and hazard damage.

\subsection{Hazard-Conditional Adaptation System}
\label{sec:adaptation}

The model adds a hazard-conditional adaptation mechanism on top of otherwise fixed operating rules. Wage setting, pricing, inventory planning, liquidity buffering, and productive-capital investment remain structural rules. Adaptation is isolated in a separate continuity-capacity stock $C_t \in [0,1]$, interpreted as generic preparedness rather than as productive capital or liquidity:

\begin{equation}
C_{t+1} = \min \left(1, (1-\delta_C) C_t + I^C_t \right)
\end{equation}

where $\delta_C$ is the continuity-capacity depreciation rate and $I^C_t$ is the firm's continuity investment. Higher continuity capacity enables one of two reported adaptation strategies, selected at the population level for scenario comparison. Both strategies share the same continuity-capacity investment rule, hazard-signal processing, and funding timing. This is a deliberate reduced-form control: the common accumulation rule holds perceived risk, budget competition, and timing fixed so that the comparison isolates how the deployment channel itself changes system outcomes. The specification should therefore not be read as claiming that supplier diversification and asset hardening have identical real-world lead times, contracting frictions, or accounting treatment.

\textbf{Hazard signals}. Each firm maintains exponentially weighted moving averages (EWMAs) of expected hazard-induced operating shortfall and nearby observed operating shortfall, using smoothing parameter 0.2 so that recent shocks receive more weight than older ones while the state variable remains compact over long runs. Direct-loss and supplier-disruption states are retained as supporting diagnostics. The nearby observed-shortfall state is constructed from hazard-related operating gaps among firms within a fixed spatial radius of 4 grid cells (1 geographic degree), allowing a firm to update its perceived continuity environment before it is directly hit itself.

\textbf{Perceived risk and target rule}. Every four steps (one year at quarterly resolution), each firm computes a continuity-risk signal as the larger of its own expected operating shortfall and the nearby observed operating shortfall:
\begin{equation}
R_{i,t} = \max \left(\hat{g}_{i,t}, \hat{g}^{\,\text{local}}_{i,t}\right),
\end{equation}
where $\hat{g}_{i,t}$ is the firm's EWMA of realized hazard-induced operating shortfall and $\hat{g}^{\,\text{local}}_{i,t}$ is the corresponding local observation. Firms annualize this per-step signal before translating it into a continuity target:
\begin{equation}
C^*_{i,t} = \min \left(1,\kappa_i \tilde{R}_{i,t}\right),
\end{equation}
where $\tilde{R}_{i,t}$ is the annualized perceived risk and $\kappa_i$ is a firm-specific continuity-sensitivity parameter drawn independently at initialization from a uniform distribution over the prescribed adaptation range, creating cross-firm heterogeneity in perceived-risk responsiveness. The comparison results shown here use $\kappa_i \sim U(0.8,1.4)$ under backup-supplier search and $\kappa_i \sim U(0.5,1.5)$ under capital hardening, as selected in the sensitivity analysis in section \ref{subsec:sensitivity}. The firm then partially closes the gap to the target:
\begin{equation}
I^C_{i,t} = \min \left(\bar{I}^C,\max(0,C^*_{i,t} - C_{i,t})\right),
\end{equation}
where $\bar{I}^C$ is the maximum continuity increment allowed at each decision update; in the reported experiments, $\bar{I}^C = 0.25$.

\textbf{Adaptation strategies}. The accumulated continuity capacity $C_{i,t}$ is a reduced-form preparedness stock that can be deployed through one of two reported strategies:

\textit{Backup-supplier search}. When a firm's primary suppliers are disrupted and residual input needs remain, the firm searches for non-primary suppliers in the same sectors that have available inventory. Backup purchases use the same market transaction function as primary purchases: real cash is transferred to the seller and real inventory is transferred to the buyer. This strategy therefore represents supply-chain diversification under stress while preserving monetary and inventory accounting.

\textit{Capital hardening}. Under this deployment channel, continuity capacity is interpreted as direct-loss preparedness such as flood-proofing, protective equipment, and continuity arrangements that reduce the common hazard damage ratio governing all three modeled direct-damage channels: capital stock, inventories, and productivity:
\begin{equation}
\ell^{\text{eff}}_{i,t} = \ell^{\text{raw}}_{i,t} \cdot (1 - C_{i,t}),
\end{equation}
where $\ell^{\text{raw}}_{i,t}$ is the raw loss fraction from the JRC damage function and $\ell^{\text{eff}}_{i,t}$ is the effective loss applied after hardening. This loss fraction is distinct from the productivity state $\phi$ in Equation \ref{eq:production}: $\ell$ is the current-period direct-loss share implied by the hazard, whereas $\phi$ is the remaining productivity multiplier after new losses are applied and subsequent recovery unfolds. This strategy represents physical hardening of productive assets.

\textbf{Funding timing}. The continuity target is evaluated before the current hazard realization, but maintenance and new continuity spending are funded only when the period closes. Continuity therefore competes with dividends and residual retained earnings rather than with same-period payroll and procurement, and newly installed continuity capacity affects the following period rather than the current one.

\textbf{Firm reorganization}. Bankrupt firms (money below the survival threshold of 1.0 monetary unit) are evaluated for in-place reorganization at 10-step global intervals rather than being replaced immediately when failure occurs. Because failures can occur between sweeps, the realized re-entry lag ranges from immediate replacement at the next sweep to at most 10 quarters under the quarterly time step, representing variable replacement times. The establishment shell, location, capital stock, inventories, and links remain in the model, while the reorganized firm inherits the adaptation state and continuity sensitivity of a successful same-sector parent. If working capital is insufficient, the firm is recapitalized through a proportional transfer from aggregate household cash holdings, so money remains inside the closed economy. The model does not track household-specific equity stakes in reorganized firms; subsequent firm payouts continue to be distributed evenly across the household sector.

\subsection{Scenario Design, Internal Validation, Ensembles, and Reproducibility}
\label{sec:ensembles}

The main comparisons are baseline without adaptation, hazard without adaptation, hazard with backup-supplier search, and hazard with capital hardening. Because the no-hazard economy exhibits a startup transient, all scenarios share an explicit 20-year no-hazard warm-up period covering 2000 Q1--2019 Q4 before the flood rasters become active. The three available Aqueduct flood rasters are then mapped onto 2020 Q1--2039 Q4 (2030 file), 2040 Q1--2059 Q4 (2050 file), and 2060 Q1--2099 Q4 (2080 file).

As a framework contribution, the model evaluation presented here combines continuously monitored diagnostics with targeted reported checks. Aggregate household-plus-firm money is tracked each step as an accounting diagnostic to verify stock-flow closure, and matched-seed ensembles control stochastic variation across scenario comparisons. A no-hazard control with adaptation enabled is also used as a dormancy check: because the continuity module is hazard-conditional, continuity targets and spending should remain inactive when direct flood losses and observed hazard shortfalls are absent. In addition, the shared no-hazard warm-up ensures that scenario comparisons are made after the startup transient has dissipated rather than during initialization artefacts, while the continuity-sensitivity analysis reported in Section \ref{subsec:sensitivity} is used to characterize how intervention intensity changes outcomes and to select strategy-specific near-efficient ranges for the final comparison. Table \ref{tab:targeted_checks} summarizes the key targeted checks and fixed parameter classes.

\begin{table}[!t]
\centering
\small
\begin{tabular}{p{2.7cm} p{4.0cm} p{3.0cm} p{4.4cm}}
\toprule
\textbf{Check / parameter class} & \textbf{Variants considered} & \textbf{Role in current paper} & \textbf{Outcome / treatment} \\
\midrule
Warm-up convergence & Shared 20-year no-hazard warm-up; convergence diagnosed on the 20-seed baseline ensemble & Validate that scenario comparison begins after startup transients & Figure \ref{fig:warmup_convergence} shows production and consumption settling before hazard onset while money drift remains near machine zero \\
Continuity sensitivity & Backup suppliers: $[0.2,0.8]$, $[0.8,1.4]$, $[1.4,2.0]$; capital hardening: $[0.5,1.5]$, $[1.5,3.0]$, $[3.0,4.5]$ & Main behavioural uncertainty governing continuity-target formation & Figure \ref{fig:sensitivity} is used to select near-efficient ranges for the final 20-seed scenario comparison \\
Fixed structural rules & Leontief coefficients, wage and price adjustment rates, liquidity-buffer rules, distance-cost draws, depreciation rules, and reorganization timing (10-step global sweep; realized lag 0--10 quarters) & Held fixed as structural stylized rules in the current framework paper & Reported as modelling assumptions rather than globally swept; broader calibration and global sensitivity remain future work \\
\bottomrule
\end{tabular}
\caption{Targeted reported checks and fixed parameter classes. Routine diagnostics such as stock-flow closure and matched-seed ensemble comparison are described in the text; the table highlights the additional convergence and sensitivity checks together with the structural rules held fixed rather than globally swept.}
\label{tab:targeted_checks}
\end{table}

\begin{figure}[!t]
\centering
\includegraphics[width=0.95\textwidth]{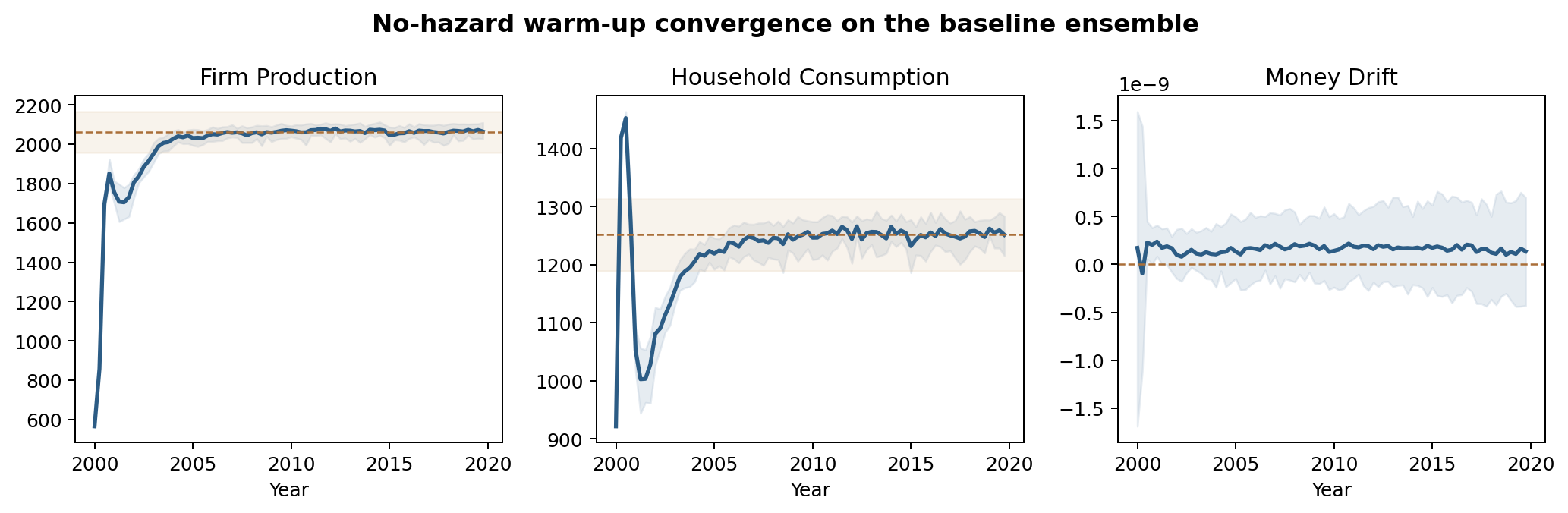}
\caption{No-hazard warm-up convergence on the 20-seed baseline ensemble. Solid lines denote ensemble means and the blue shaded region denotes the 10th--90th percentile range. For production and consumption, the dashed line and tan shaded band mark the late-warm-up mean over 2015--2019 together with a $\pm 5\%$ tolerance band. Production remains within that band from 2003 Q2 onward and consumption from 2003 Q4 onward. For money drift, the dashed line marks zero and the only shaded band is the time-varying 10th--90th percentile range; drift remains at machine-zero scale throughout.}
\label{fig:warmup_convergence}
\end{figure}

Figure \ref{fig:warmup_convergence} reports the no-hazard warm-up convergence diagnostic computed from the 20-seed baseline ensemble. Aggregate firm production remains within 5\% of its late-warm-up mean from 2003 Q2 onward and household consumption from 2003 Q4 onward. Money drift is plotted against a zero benchmark and its ensemble spread remains at machine-zero scale throughout the warm-up, with absolute drift staying below $1.93 \times 10^{-9}$. The shared 2000 Q1--2019 Q4 burn-in is therefore conservative relative to the observed convergence time.

Model outcomes are stochastic because flood realizations, labour-market matching, and within-tier firm ordering all depend on random draws. Scenario comparisons are therefore evaluated as ensembles of independent runs that differ only by random seed. The presented results are configured as a matched 20-seed ensemble for each of the four substantive scenarios with ensemble means together with 10th--90th percentile bands, using the strategy-specific continuity-sensitivity ranges selected in Section \ref{subsec:sensitivity}: $[0.8,1.4]$ for backup-supplier search and $[0.5,1.5]$ for capital hardening.

\subsection{Systemic Cascade Diagnostics}
\label{sec:cascade}

To isolate systemic transmission beyond directly flooded firms, the model also records a persistent direct-exposure state for each firm. A firm is classified as ``never hit'' at time $t$ if it has not experienced any direct flood loss up to that point in the simulation. From this classification, the framework reports four hazard-only diagnostics: (i) the cumulative share of firms ever directly hit, (ii) the share of firms that remain never hit but are currently experiencing supplier disruption, (iii) the share of total supplier-disruption burden borne by firms that remain never hit, and (iv) the share of aggregate production generated by firms that remain never hit. These diagnostics are computed directly during the simulation run and exported with the main model summary.

\section{Illustrative Application Results}
\label{sec:results}

\begin{figure}[!t]
\centering
\includegraphics[width=0.95\textwidth]{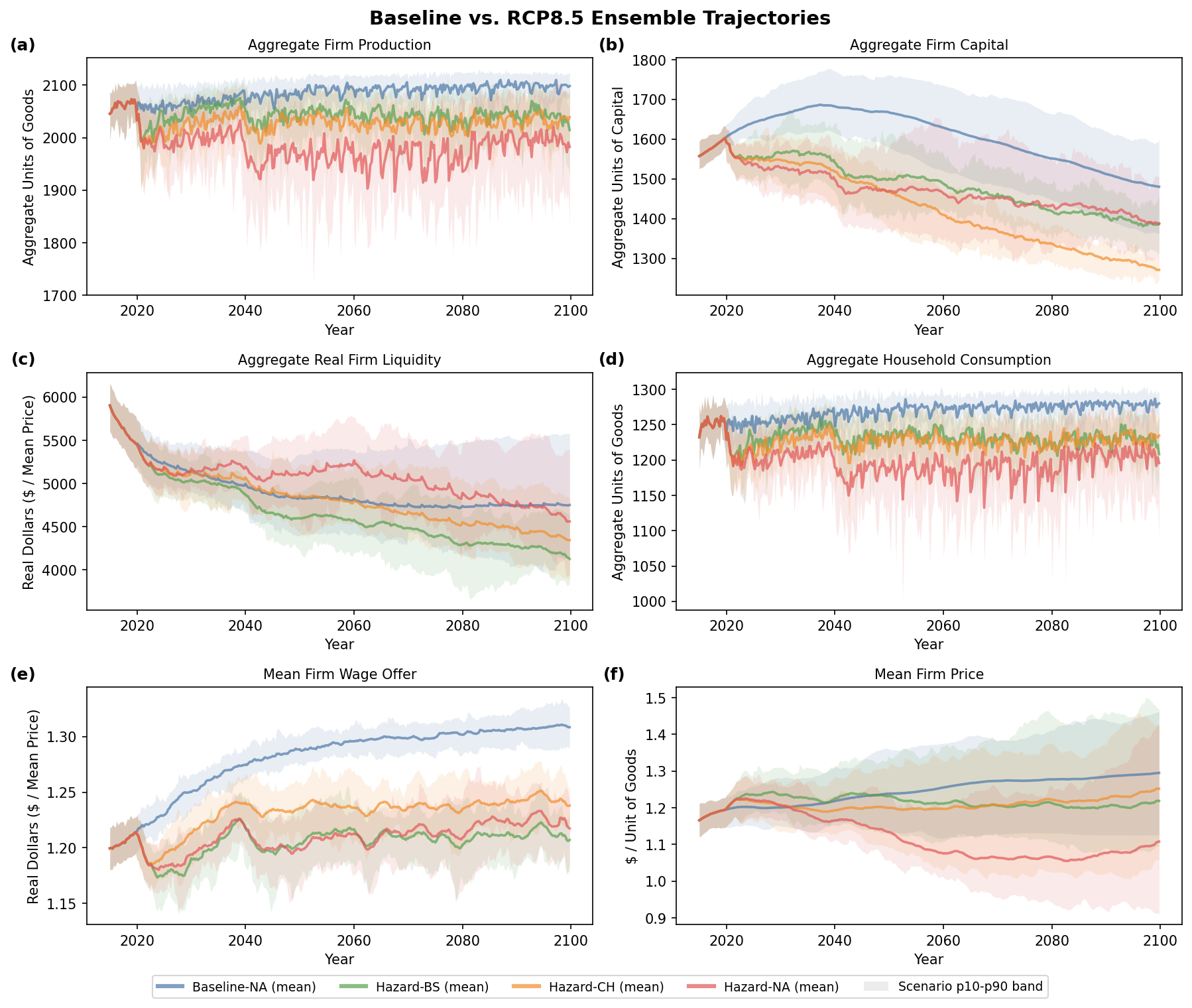}
\caption{Ensemble trajectories for the four reported scenarios. Panels show (a) aggregate firm production, (b) aggregate firm capital, (c) aggregate real firm liquidity (deflated by mean price), (d) aggregate household consumption, (e) mean real wage offer (deflated by mean price), and (f) mean firm price. Scenario abbreviations use NA for no adaptation, BS for backup-supplier search, and CH for capital hardening. Lines denote ensemble means and the shaded region denotes the scenario-specific 10th--90th percentile range.}
\label{fig:ts}
\end{figure}

The illustrative economic network is run under RCP8.5 riverine flooding from 2020 Q1 to 2099 Q4 after the shared 2000 Q1--2019 Q4 warm-up, with quarterly time steps. The main scenario comparison is reported as a matched 20-seed ensemble using the workflow described in Section \ref{sec:ensembles}. Figure \ref{fig:ts} reports the main macro comparison: one baseline benchmark, one hazard benchmark without continuity investment, and two hazard-with-adaptation cases (backup-supplier search and capital hardening), each summarized by an ensemble mean together with a 10th--90th percentile band. These results should be read as an internal evaluation of the framework under a stylized network configuration, not as an externally validated forecast for a specific real economy.

The final ensemble confirms that repeated flood shocks materially weaken the sample economy. Over the last 40 recorded quarters (2090 Q1--2099 Q4), hazard without adaptation lowers production by 4.9\%, consumption by 5.6\%, capital by 6.0\%, and real wages by 6.4\% relative to the no-hazard baseline. Mean prices are 15.7\% lower under hazard without adaptation than under the baseline, indicating that part of the apparent nominal relief in the stressed economy comes from a contraction in demand rather than from successful resilience. Both adaptation strategies then raise prices relative to hazard without adaptation, but even in the final decade their mean price levels remain below the no-hazard baseline, by 4.1\% under capital hardening and 6.3\% under backup-supplier search.

\subsection{Adaptation Activation}
\label{subsec:production}

The baseline settles into a stable no-hazard operating regime after the warm-up, while the baseline and hazard no-adaptation controls keep continuity metrics at zero by construction. Average continuity capacity, perceived continuity risk, and continuity targets therefore remain zero in those control runs. Under repeated floods, however, continuity capacity rises to 0.163 under capital hardening and 0.209 under backup-supplier search in the final decade, with corresponding continuity targets of 0.171 and 0.223. The higher continuity-capacity stock under backup-supplier search reflects a stronger supplier-disruption signal and shows that the adaptation mechanism is endogenously activating under hazard stress rather than being mechanically imposed.

\subsection{Macro Outcomes}
\label{subsec:capital}

Relative to hazard without adaptation, both adaptation strategies recover about 2\% of lost production and consumption, but they do so through distinct channels and with different financial-side consequences. Capital hardening raises final-decade production by 1.9\% and consumption by 1.9\% relative to hazard without adaptation, while lowering realized direct loss from 0.0397 to 0.0293 (-26.1\%). It also raises the real wage by 1.6\%. However, mean prices rise by 13.8\%, aggregate capital falls by 8.3\%, and real firm liquidity falls by 5.8\% relative to hazard without adaptation. Backup-supplier search raises production by 2.4\% and consumption by 1.9\%, leaves aggregate capital roughly unchanged relative to hazard without adaptation (-0.9\%), and lowers supplier disruption from 0.0119 to 0.0062 (-47.7\%). But realized direct loss falls by only 5.3\%, real wages fall by 0.9\%, and real firm liquidity falls by 9.8\%. Relative to the no-hazard baseline, both adaptation scenarios still remain below baseline production and consumption levels, so the substantive result is partial stabilization rather than full recovery. Thus capital hardening is the stronger direct-protection mechanism, whereas backup-supplier search is the stronger indirect continuity mechanism; neither dominates across all macro and financial margins.

\subsection{Cascade Diagnostics}
\label{subsec:wages}

\begin{figure}[!t]
\centering
\includegraphics[width=0.95\textwidth]{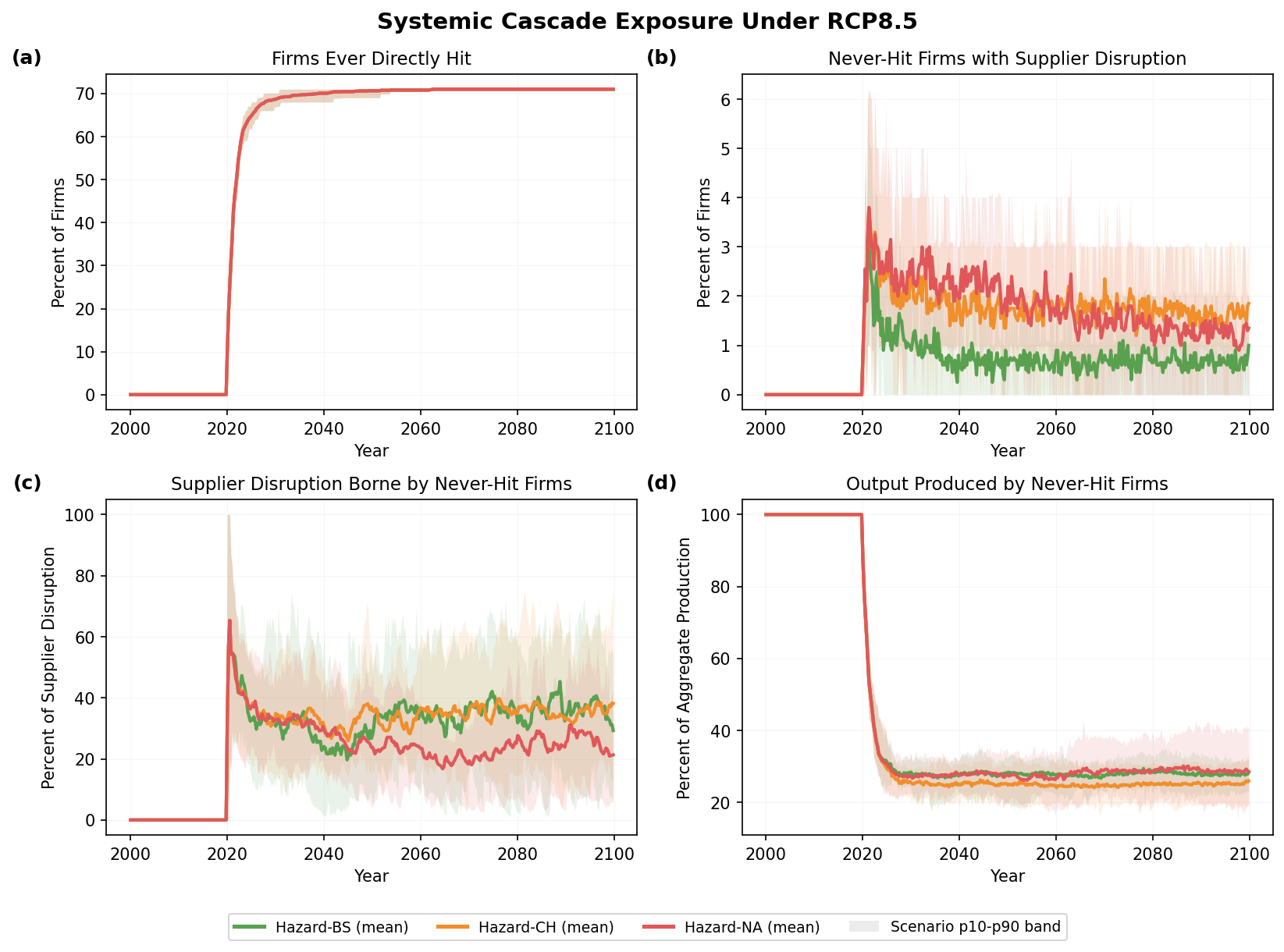}
\caption{Hazard-only systemic cascade diagnostics under RCP8.5. Scenario abbreviations use NA for no adaptation, BS for backup-supplier search, and CH for capital hardening. Panel (a) shows the cumulative share of firms ever directly hit. Panel (b) shows the share of all firms that remain never directly hit but are currently experiencing supplier disruption. Panel (c) shows the share of total supplier-disruption burden borne by firms that remain never directly hit. Panel (d) shows the share of aggregate production generated by firms that remain never directly hit. Lines denote ensemble means and the shaded region denotes the scenario-specific 10th--90th percentile range.}
\label{fig:cascade_risk}
\end{figure}

Figure \ref{fig:cascade_risk} reports the hazard-only cascade diagnostics. By the end of the simulation (2099 Q4) roughly 71\% of firms have been directly hit at least once in all three hazard scenarios, leaving about 29\% of firms never directly flooded. Despite never being directly hit, this subset still produces 25--29\% of aggregate output in the final decade and continues to bear a non-trivial share of supply-chain stress. Averaged over the final decade (2090 Q1--2099 Q4), the share of all firms that are both never hit and currently disrupted is 1.26\% under hazard without adaptation. Backup-supplier search lowers that share to 0.68\%, whereas capital hardening raises it to 1.61\%. The share of total supplier-disruption burden borne by never-hit firms, however, rises from 26.2\% without adaptation to about 35--36\% under both adaptation cases. The two cascade metrics capture different aspects of indirect risk. Under backup-supplier search, both the current disruption incidence on never-hit firms and their absolute disruption burden fall, even though the remaining disruption becomes more concentrated on never-hit firms as directly hit firms are relieved more strongly. Under capital hardening, by contrast, the current never-hit disruption share and the relative burden share both rise, indicating weaker protection against indirect cascades despite lower direct loss.

\subsection{Sensitivity Analysis}
\label{subsec:sensitivity}

\begin{figure}[!t]
\centering
\includegraphics[width=\textwidth]{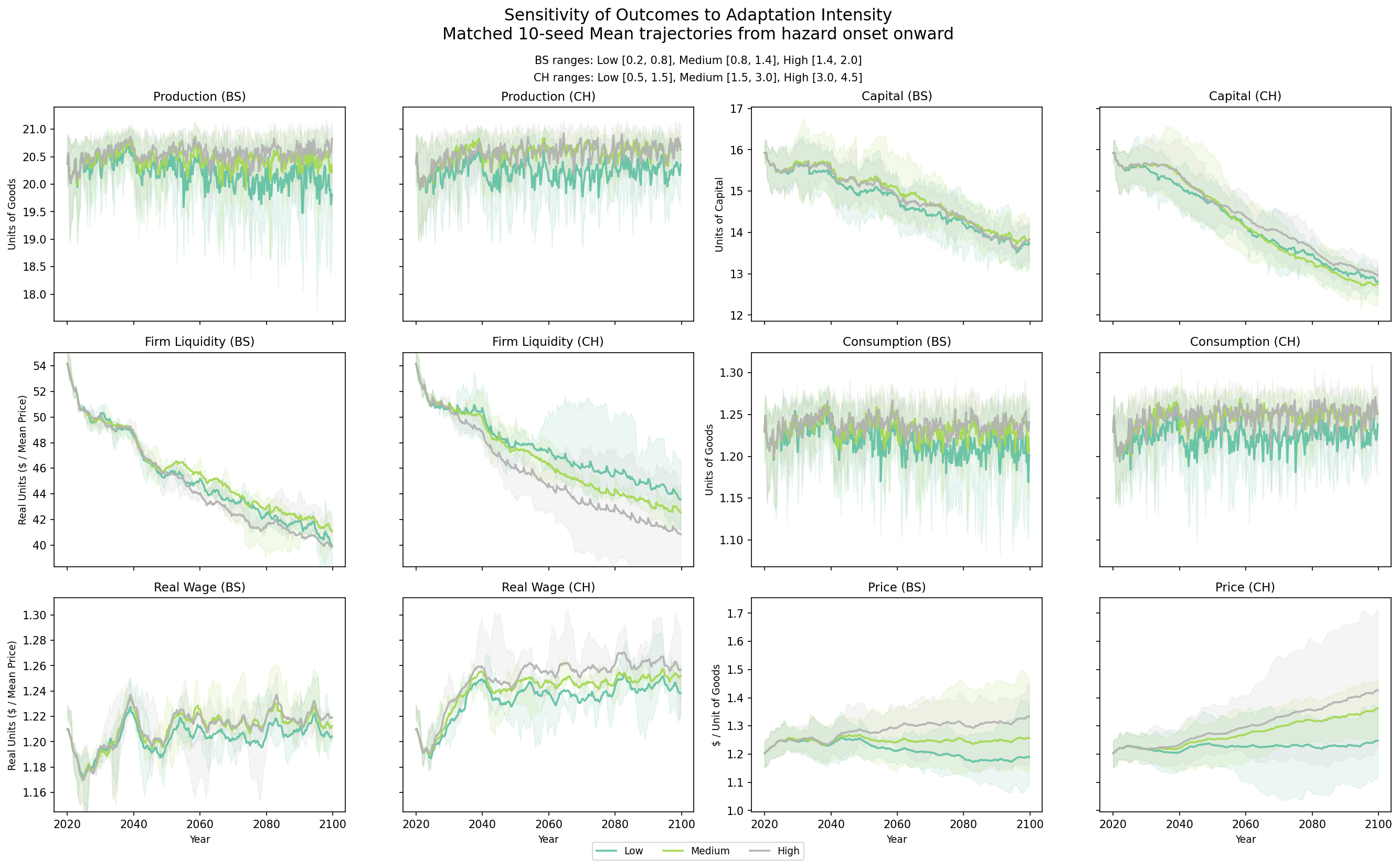}
\caption{Combined matched-seed ensemble sensitivity analysis for the two reported adaptation strategies, plotted from hazard onset in 2020 Q1 after the shared warm-up. Backup-supplier search (BS) and capital hardening (CH) are shown side-by-side for each metric: production and capital in the top row, real liquidity and consumption in the middle row, and real wage and price in the bottom row. BS ranges are Low $[0.2,0.8]$, Medium $[0.8,1.4]$, and High $[1.4,2.0]$; CH ranges are Low $[0.5,1.5]$, Medium $[1.5,3.0]$, and High $[3.0,4.5]$. The final 20-seed comparison uses $[0.8,1.4]$ for backup-supplier search and $[0.5,1.5]$ for capital hardening.}
\label{fig:sensitivity}
\end{figure}

The two reported adaptation strategies share one main behavioural parameter: the continuity-sensitivity coefficient that maps perceived hazard risk into a continuity target. Figure \ref{fig:sensitivity} reports matched-seed ensemble sensitivity analysis across several ranges of this parameter, run separately for backup-supplier search and capital hardening. The purpose of the exercise is not to exhaustively sensitivity-test every model coefficient, but to characterize how intervention intensity changes outcomes and to select strategy-specific near-efficient ranges for the final 20-seed scenario comparison. The tested low/medium/high ranges are strategy-specific exploratory brackets chosen from preliminary matched-seed runs to span weak, moderate, and strong continuity responses without overlap within each strategy. For backup-supplier search the brackets are $[0.2,0.8]$, $[0.8,1.4]$, and $[1.4,2.0]$; for capital hardening they are $[0.5,1.5]$, $[1.5,3.0]$, and $[3.0,4.5]$. Because the continuity-sensitivity coefficient operates through different deployment channels across the two strategies, these labels are not intended to represent equal intervention intensities across strategies.

The sensitivity outputs should therefore be interpreted as strategy-selection diagnostics rather than as evidence that the continuity-sensitivity parameter is immaterial. For capital hardening, the low range $[0.5,1.5]$ yields nearly the same production and consumption gains as the stronger ranges while materially limiting inflation and preserving more real liquidity, so it is selected for the final ensemble comparison. For backup-supplier search, the medium range $[0.8,1.4]$ is selected. Relative to the low range, it improves production, consumption, capital, real liquidity, and supplier-disruption relief, with only a moderate additional price increase; the high range adds further inflation with little extra macro gain.

\section{Discussion}
\label{sec:discussion}

The final 20-seed ensembles partially support the hypothesis that hazard-conditional continuity adaptation can improve outcomes relative to hazard without adaptation while separating direct-loss mitigation from indirect continuity preservation. Both reported strategies recover about 2\% of lost production and consumption. Capital hardening lowers direct flood losses most strongly, whereas backup-supplier search lowers supplier disruption most strongly. Yet neither restores baseline conditions, and both reduce real firm liquidity relative to hazard without adaptation because preserving activity in a supply-constrained economy raises the price level. Taken together, Figures \ref{fig:ts} and \ref{fig:cascade_risk} imply a specific interpretation of hazard-conditional adaptation in this closed economy: lower prices under hazard without adaptation are not a sign of successful stabilization, but largely a symptom of demand-side contraction. Both adaptation strategies prevent part of that collapse, so prices rise as more activity is sustained in a supply-constrained system, even though the adapted hazard scenarios still remain below the no-hazard baseline price level. Capital hardening buys direct-loss protection but does not preserve capital in this configuration, showing that direct-loss protection does not guarantee either higher capital or higher real liquidity. Backup-supplier search buys stronger indirect continuity, but it does so with the lowest real firm liquidity of the three hazard scenarios. The adapted price paths should therefore be read as a partial normalization from the depressed no-adaptation state rather than as an overshoot beyond the no-hazard benchmark.

The broader contribution is a controlled software workflow for tracing how acute hazards propagate through a supply-chain economy under explicit accounting rules and reproducible scenario configuration. The internal evaluation reported in this paper shows stock-flow closure, dormant adaptation in no-hazard controls, conservative warm-up convergence before hazard onset, and stable matched-seed ensemble separation across scenarios. These checks do not amount to external calibration against a specific regional economy, but they do establish that the framework behaves coherently enough for methodological experiments on acute hazard propagation and firm adaptation.

The two reported adaptation channels then clarify that resilience is not one-dimensional. Backup-supplier search protects production flows by reducing supplier disruption, whereas capital hardening protects assets by lowering realized direct loss. Because both strategies are funded from residual cash, each also trades off against retained liquidity, and neither restores the no-hazard baseline. The model therefore distinguishes among protecting flows, protecting assets, and preserving real liquidity rather than collapsing them into a single resilience score.

The cascade diagnostics make this distinction visible. Rather than inferring systemic risk only from aggregate macro trajectories, the reporting workflow quantifies how much disruption is borne by firms that remain never directly flooded. In final-decade averages, roughly 29\% of firms remain never hit while still bearing 26--36\% of the supplier-disruption burden and generating roughly one quarter of output. The hazard-only cascade figure should therefore be read as a decomposition of the aggregate losses shown in Figure \ref{fig:ts}, not as a separate welfare ranking. Under backup-supplier search, the share of all firms that are both never hit and currently disrupted falls, and the absolute disruption burden on never-hit firms also falls, even though their share of total disruption rises because disruption declines more strongly on the directly hit subset. Under capital hardening, by contrast, the current never-hit disruption share and the burden share both rise, showing that lower direct loss does not by itself imply weaker indirect cascades.

Importantly, the acute climate shocks are sampled stochastically per grid cell based on the associated return period. Only a subset of firms are directly subject to acute shocks at any given time step. Nevertheless, supply-chain disruptions propagate through the network, creating input bottlenecks for firms without direct hazard exposure. Traditional climate risk assessment approaches that consider only direct damages to assets would underestimate such systemic risks. While some approaches consider indirect damages as a result of operational disruptions, these are still direct exposures to hazards and do not capture systemic risks without any direct exposure to acute events such as flooding.

While top-down econometric damage functions do attempt to capture some systemic risks, they still face the extrapolation and functional-form limitations discussed in Section \ref{sec:intro}. The bottom-up agent-based approach used here addresses these limitations in three ways. First, regarding \textbf{extrapolation to new regimes}: the ABM generates emergent economic behaviours under climate conditions far outside historical experience. The model captures path dependencies in continuity capacity, operating-shortfall expectations, and supplier disruption that polynomial damage functions would miss. Second, regarding \textbf{limited functional forms}: the hazard-conditional adaptation mechanism makes firm responses explicit through agent-level decision rules that update continuity targets as perceived hazard conditions worsen. This generates heterogeneous, state-dependent responses across agents based on exposure, nearby observed shortfalls, sector, and network position. Third, regarding \textbf{cascading effects}: the supply-chain structure creates propagation pathways for climate shocks that extend far beyond directly exposed assets. Firms farther downstream in the network can experience production constraints when upstream suppliers are damaged, even if the downstream firm itself faces no direct hazard exposure.

\subsection{Limitations}
\label{subsec:limitations}

This study has several limitations. Most importantly, we use an illustrative economic network of 100 firms across the globe as a proof of concept. The outcomes observed are therefore illustrative examples within this sample network and do not represent predictions of actual economy-wide impacts. The model is not externally calibrated or validated against a specific historical flood, observed production network, or firm-level recovery dataset. For more actionable quantitative conclusions, future work should consider firm-level or sector-level input-output data to calibrate trade relationships. Historical data across firms would also enable a more fine-tuned approach for quantifying the distribution of firm behaviours under normal and acute stress scenarios.

The scope of the current study is limited to acute riverine flooding under RCP8.5 using the JRC damage function. A broader scope of acute and chronic hazards under additional scenarios, as well as incorporation of alternative damage-function frameworks, would enable more complete risk assessment. The current continuity-capacity abstraction also deliberately suppresses strategy-specific lead times, contracting frictions, and accounting distinctions between operational diversification and physical hardening; future work could relax that reduced-form control once strategy-specific empirical data are available.

\section{Conclusions}
\label{sec:conclusions}

We presented an open-source spatial agent-based framework that integrates acute physical climate risk with hazard-conditional continuity adaptation to capture how economic agents respond to new risk regimes. We used Aqueduct riverine flooding under RCP8.5 and JRC region-specific damage functions to illustrate this approach on an idealized economic network of 100 commodity, manufacturing, and retail firms with 1000 households distributed across the globe at quarterly temporal resolution.

Key design choices are final-goods-only household demand, demand-driven firm planning, labour-consistent startup seeding, bounded working-capital finance for current operations, retained-earnings capital formation, phased within-period markets, stock-flow-consistent continuity spending, and ensemble-based scenario reporting. Within this structure, firms update continuity targets from adaptive expectations of worsening hazard-induced operating shortfall and nearby observed disruptions, and deploy accumulated continuity capacity through one of two reported strategies: backup-supplier search or capital hardening.

The final 20-seed ensembles establish a coherent internal scenario comparison between a no-hazard baseline, hazard without adaptation, hazard with backup-supplier search, and hazard with capital hardening, while also decomposing how much stress is borne by firms that remain never directly flooded. Hazard without adaptation reduces production and consumption relative to baseline while also depressing prices through demand contraction. Capital hardening lowers realized direct losses most strongly, whereas backup-supplier search lowers supplier disruption most strongly; both recover modest amounts of output and consumption, but neither restores baseline real liquidity. Together with the stock-flow, no-hazard control, warm-up, ensemble, and sensitivity diagnostics, these results show that the framework provides a disciplined basis for integrating acute hazard data, damage functions, and systemic economic propagation in one reproducible workflow. The framework therefore provides a concrete basis for assessing both direct climate damages and the systemic resilience value of firm-level adaptation in supply-chain economies, while showing that protecting assets, protecting flows, and preserving real liquidity need not coincide. More broadly, it enables quantification of indirect system risks from acute hazards, including disruption borne by firms that are never directly flooded, which remains largely missing from mainstream direct-damage climate-risk workflows.

\section*{CRediT authorship contribution statement}
\textbf{Yara Mohajerani}: Conceptualization, Methodology, Software, Validation, Formal analysis, Investigation, Resources, Data curation, Writing -- original draft, Writing -- review \& editing, Visualization.

\section*{Declaration of competing interest}
The author, Yara Mohajerani, is the founder and president of Quantile Labs Inc.

\section*{Software and data availability statement}
\textbf{Software.} The \textit{Spatial Climate-Economy Agent-Based Model} was developed by Yara Mohajerani at Quantile Labs Inc. and first made publicly available in 2025. It is written in Python 3.11+, distributed free of charge under the BSD 3-Clause License, and hosted in a public GitHub repository (\href{https://github.com/yaramohajerani/spatial-climate-ABM}{project repository}). The core source and configuration files are lightweight (less than 1 MB excluding hazard rasters, figures, and generated outputs) and depend primarily on Mesa, NumPy, pandas, rasterio, GeoPandas, matplotlib, and openpyxl. The reported 100-firm experiments run on standard desktop or laptop hardware; because hazard rasters are sampled lazily at agent locations, the illustrative experiments do not require loading full global rasters into memory.

\textbf{Data.} Riverine flood projections are obtained from the World Resources Institute Aqueduct database \citep{ward2020aqueduct}, which distributes the hazard rasters online. Damage functions are taken from the JRC Global Flood Depth-Damage Functions \citep{huizinga2017global}. The repository contains the parameter files, topology files, plotting scripts, and example summary/member CSV outputs used to reproduce the workflow reported here. Reproducibility is supported through fixed random seeds and \texttt{Meta\_*} metadata fields written to the outputs.

\section*{Acknowledgements}
This work was supported by Quantile Labs Inc.

\medskip
\small
\bibliographystyle{unsrtnat}
\bibliography{bibliography}  

\end{document}